\documentclass[pdflatex,sn-basic]{sn-jnl}% Basic Springer Nature Reference Style/Chemistry Reference Style

\jyear{2021}%

\theoremstyle{thmstyleone}%
\theoremstyle{thmstyletwo}%

\theoremstyle{thmstylethree}%

\usepackage{multirow}
\usepackage{array}
\usepackage{amsmath,amssymb,amsfonts}
\usepackage{csquotes}
\usepackage{xurl}
\usepackage{float}
\usepackage{color}

\begin{document}
	
	\title[BT-Unet framework for biomedical image segmentation]{BT-Unet: A self-supervised learning framework for biomedical image segmentation using Barlow Twins with U-Net models}
	
	
	\author*[1]{\fnm{Narinder Singh} \sur{Punn}}\email{pse2017002@iiita.ac.in}
	
	\author[1]{\fnm{Sonali} \sur{Agarwal}}\email{sonali@iiita.ac.in}
	
	\affil[1]{\orgdiv{Dept. of information Technology}, \orgname{Indian Institute of Information Technology Allahabad}, \orgaddress{\city{Prayagraj}, \state{Uttar Pradesh}, \country{India}}}

	
	\abstract{Deep learning has brought the most profound contribution towards biomedical image segmentation to automate the process of delineation in medical imaging. To accomplish such task, the models are required to be trained using huge amount of annotated or labelled data that highlights the region of interest with a binary mask. However, efficient generation of the annotations for such huge data requires expert biomedical analysts and extensive manual effort. It is a tedious and expensive task, while also being vulnerable to human error. To address this problem, a self-supervised learning framework, BT-Unet is proposed that uses the Barlow Twins approach to pre-train the encoder of a U-Net model via redundancy reduction in an unsupervised manner to learn data representation. Later, complete network is fine-tuned to perform actual segmentation. The BT-Unet framework can be trained with a limited number of annotated samples while having high number of unannotated samples, which is mostly the case in real-world problems. This framework is validated over multiple U-Net models over diverse datasets by generating scenarios of a limited number of labelled samples using standard evaluation metrics. With exhaustive experiment trials, it is observed that the BT-Unet framework enhances the performance of the U-Net models with significant margin under such circumstances.}

	
	
	
	\maketitle
	
	\section{Introduction}\label{sec1}
	With the advent of advancements in deep learning technologies, there is a significant gain in the momentum of its applications in biomedical image analysis such as classification, localization, segmentation, etc.~\citep{ker2017deep}. Most medical applications require segregating the objects or regions (damaged tissues, cells, nuclei, organs, etc.) with fine boundaries using medical imaging such as CAT scans, X-Rays, Ultrasound, etc. for diagnosis, monitoring and treatment. This delineation is generally performed by expert clinicians or radiologists which is a complex and tedious task. With biomedical image segmentation being a precursor to computer-aided classification/localization, various deep learning based approaches are developed to automate the segmentation process for faster diagnosis and better treatment~\citep{haque2020deep}. Among these approaches, U-Net~\citep{ronneberger2015u} based segmentation models gained significant popularity due to its mutable and modular structure that would result in the state-of-the-art diagnosis system~\citep{punn2021modality}.
	
	However, such potential of deep learning segmentation models is only unlocked by training the models with a large amount of annotated data i.e., a fully supervised approach. Moreover, efficient generation of the annotations for such huge data requires expert biomedical analysts and extensive manual effort. It is a tedious and expensive task, while also being vulnerable to human error. To address this issue, various strategies are adopted to efficiently train the model with limited labelled data samples such as data augmentation, transfer learning, self-supervised learning, etc. In image data augmentation~\citep{shorten2019survey} the aim is to increase the number of labelled data by geometric transformations, feature space augmentation, generative adversarial networks, etc. However, the diversity of the augmented samples is limited by the available annotated samples which could cause an overfitting problem in the model. Several attempts are also made towards transfer learning to alleviate the performance of model with limited annotated data samples. Though this strategy works very well with natural images, but is ineffectual in biomedical image analysis~\citep{alzubaidi2020towards, raghu2019transfusion} due to large variation in the associated complex patterns of biomedical imaging as compared to natural images. 
	
	Self-supervised learning~\citep{jing2020self} is an emerging technology that is effectively closing the gap with fully supervised methods on large computer vision benchmarks. It provides an effective solution to the limited availability of annotated data. Here, the aim is to perform pre-training with an unsupervised strategy for learning useful and better representations of the data samples. The pre-trained model is then fine-tuned with limited annotated samples to adopt the actual task such as segmentation, classification, etc. The recent development in self-supervised learning methods can be categorized as contrastive learning (MoCo~\citep{he2020momentum}, PIRL~\citep{misra2020self}, SimCLR~\citep{chen2020simple}), clustering (DeepCluster~\citep{caron2018deep}, SeLA~\citep{asano2019self}, SwAV~\citep{caron2020unsupervised}), distillation (BYOL~\citep{grill2020bootstrap}, SimSiam~\citep{chen2021exploring}) and redundancy reduction (Barlow Twins~\citep{zbontar2021barlow}). The approaches categorized under contrastive learning, clustering and distillations work based on the similarity maximization that requires efficient generation of the positive (related images) and negative (unrelated images) samples for pre-training. However, in biomedical image analysis identifying the negative samples is a tedious and complex task~\citep{zeng2021positional} due to similarities and dissimilarities at low and high level feature representations respectively, whereas in Barlow Twins there is no such requirement; therefore, more suitable for biomedical image segmentation. With this motivation, in the present work a self-supervised learning framework called BT-Unet is proposed for biomedical image segmentation, where Barlow Twins strategy is integrated with U-Net segmentation models. The main contributions of the present research work are highlighted below:
	
	\begin{itemize}
		\item The challenge of limited biomedical annotated data availability is addressed by integrating redundancy reduction based self-supervised learning approach with U-Net segmentation models.
		\item The pre-training of the U-Net encoder is performed with the Barlow Twins strategy to learn feature representations in an unsupervised manner (without data annotations).
		\item The effect of pre-training on biomedical image segmentation performance is analyzed with multiple U-Net models over diverse datasets.
	\end{itemize}
	
	The rest of the paper is divided into several sections, where Section \ref{sec2} presents the literature review of the recent developments in the self-supervised segmentation approaches, followed by the methods adopted in the proposed framework in Sections \ref{sec3} and \ref{sec4}. Sections \ref{sec5} and \ref{sec6} present the experimental setup and the obtained results respectively. Finally, concluding remarks are presented in Section \ref{sec7}.

	\section{Related work}\label{sec2}
	In recent years, due to developments in deep learning technologies, the researchers have developed a keen interest in computer-aided diagnosis systems to promote better healthcare services with a variety of applications~\citep{lei2020medical} such as classification, detection, segmentation, etc. With segmentation being one of the critical aspects of diagnosis and follow up treatment plans, various deep learning based segmentation models are developed. However, the use of self-supervised learning strategies to improve the segmentation performance is relatively least explored.
	
	In the context of biomedical image segmentation, most of these approaches can be grouped into pretext based and contrastive learning based strategies.	In pretext based self-supervised learning, a proxy task is performed to learn the feature representations. There are variety of pretext or proxy tasks that can be used for pre-training such as inpainting~\citep{pathak2016context}, jigsaw puzzles~\citep{noroozi2016unsupervised}, predicting the position of image patches~\citep{doersch2015unsupervised}, predicting rotations~\citep{gidaris2018unsupervised}, etc. However, there is a huge gap or variation between these tasks and the actual or downstream tasks due to which these strategies achieved limited success in deep learning applications. In contrastive learning based strategies, the feature representations are learned by effectively distinguishing the positive (similar) and negative (dissimilar) pairs. Recently, contrastive learning based unsupervised feature representations have gained significant interest. Following this context, \cite{chaitanya2020contrastive} proposed a contrastive learning framework that adapts global and local features using unannotated samples during pre-training in a stage-wise manner for biomedical image segmentation. Similarly, \cite{zheng2021hierarchical} proposed a hierarchical self-supervised framework, where multiple heterogenous datasets across multiple modalities are utilized for multi-level contrastive pre-training to adapt the multiple segmentation tasks by fine-tuning. \cite{dhere2021self} performed kidney segmentation with a self-supervision strategy where contrastive learning is used with pretext task defined as the classification of the pair of kidneys belonging to the same side, where the pre-training is performed using a siamese network and the pre-trained encoder is fine-tuned in the U-Net model for final segmentation. 
	
	However, these contrastive approaches require generating effective positive and negative pairs which is not feasible in every task such as nuclei segmentation or skin lesion segmentation, where the input samples are almost related and it is relatively hard to generate negative pair of samples~\citep{li2021imbalance}. Following this context, a redundancy reduction based strategy is adopted that does not require generation of positive and negative pairs for pre-training. Here, the aim is to obtain invariant and independent feature representations for every neuron of a model by minimizing the true and observed cross-correlation matrices which is the opposite of mutual information between the representations.

	\section{Methods}\label{sec3}
	In this section, the background knowledge of the redundancy reduction based Barlow Twins approach for self-supervised learning along with U-Net based models are presented that are integrated with Barlow Twins for biomedical image segmentation.
	
	\subsection{Barlow Twins}
\begin{figure}
		\centering
		\includegraphics[width=0.7\columnwidth] {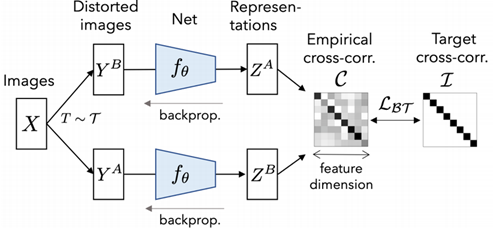}
		\caption{Schematic representation of Barlow Twins~\citep{zbontar2021barlow}.}
		\label{fig1}
	\end{figure}

	Inspired by Horace Barlow’s efficient coding hypothesis, where neurons communicate via spiking codes which aim to reduce the redundancy between neurons, \cite{zbontar2021barlow}  proposed a redundancy reduction based Barlow Twins (BT) framework for self-supervised learning. Here, the objective is to make each neuron satisfy two conditions that are producing feature representations: 1) Invariance – invariant under different augmentations, and 2) Reduce redundancy – independent of other neurons. The overall BT framework is presented in Fig.~\ref{fig1}. In this, two identical encoders ($f_\theta$, siamese net), sharing the same parameters and weights, generates feature representations ($Z^A$ and $Z^B$) of the augmented or corrupted images ($Y^A$ and $Y^B$). Later, a cross-correlation matrix ($\mathcal{C}$) is generated from batch normalized feature representations: $Z^A$ and $Z^B$. Finally, to satisfy the above two properties, the model is fine-tuned to make the matrix $\mathcal{C}$ fairly similar to an identity matrix with a loss function, $\mathcal{L}_{BT}$ defined as shown in Eq.~\ref{eq1}. In the BT-Unet framework, the encoder of the U-Net models is pre-trained with the BT strategy and later fine-tuned to perform actual segmentation.
	\begin{equation}
		\mathcal{L_{BT}} = \sum_i  (1-\mathcal{C}_{ii})^2  + ~\lambda\sum_{i}\sum_{j \neq i} {\mathcal{C}_{ij}}^2
		\label{eq1}
	\end{equation}
	\begin{equation}
		\mathcal{C}_{ij}=\frac{
			\sum_b z^A_{b,i} z^B_{b,j}}
		{\sqrt{\sum_b {(z^A_{b,i})}^2} \sqrt{\sum_b {(z^B_{b,j})}^2}}
		\label{eq2}
	\end{equation}
	where $\sum_i (1-\mathcal{C}_{ii})^2$ is an invariance term (diagonal or identity term) to direct neurons to produce same output under different augmentations, and $\sum_{i}\sum_{j \neq i} {\mathcal{C}_{ij}}^2$ is a redundancy reduction term (off-diagonal term) to make each neuron produce different output. The term $\lambda$ is used to balance the contribution of invariance and redundancy reduction terms, which however is kept equal to $0.2$~\citep{zbontar2021barlow}.
	
	\subsection{U-Net models}
\begin{figure}
		\centering
		\includegraphics[width=0.7\columnwidth] {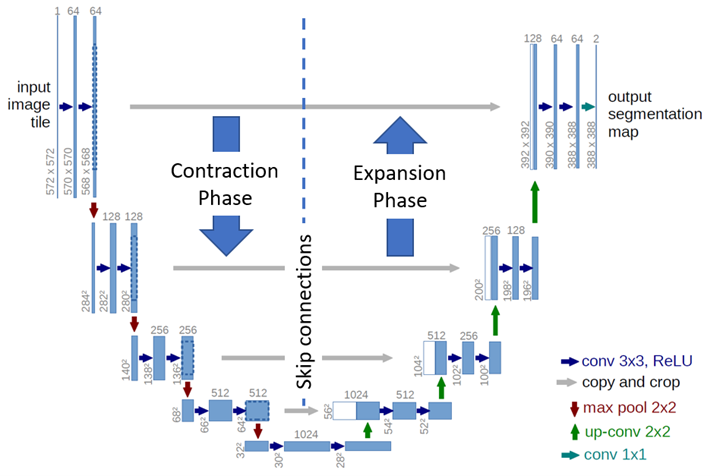}
		\caption{U-Net architecture~\citep{ronneberger2015u}.}
		\label{fig2}
	\end{figure}

	U-Net~\citep{ronneberger2015u} is the most widely used model for biomedical image segmentation. As shown in Fig.~\ref{fig2}, it follows symmetric encoder-decoder design to extract and reconstruct the feature maps respectively. The encoder phase uses the stack of ReLU activated convolution and pooling operations for feature extraction and later these feature maps are concatenated with the corresponding decoder block using the skip connections for feature up-sampling operation. Finally, $1 \times 1$ convolution is used in the output layer to generate a segmentation mask and categorize each pixel of an input image. The model was trained with the pixel-wise weighted cross-entropy function as defined in the Eq.~\ref{eq3}. The U-Net model achieved state-of-the-art results in the ISBI cell tracking challenge. With this potential of the U-Net model, various U-Net based models are developed for different biomedical image segmentation applications~\citep{punn2021modality}.
	
	\begin{equation}
		E = \sum_{{x} \in \Omega}\left(w_c({x}) + w_0 \cdot \exp\left( - \frac{(d_1({x}) + d_2({x}))^2}{2\sigma^2}\right)\right) \log({p}_{\ell({x})}({x}))
		\label{eq3}
	\end{equation}
	where $p_k(x)$ is the output softmax function, $d_1$ and $d_2$ indicate the distances to the nearest and second nearest boundary points, $w_c$ represents weight map, $w_o$ and $\sigma$ are constants.
	
	In the present article, U-Net, attention U-Net (AU-Net)~\citep{oktay2018attention}, inception U-Net (I-Unet)~\citep{punn2020inception} and residual cross spatial attention guided inception U-Net (RCA-IUnet)~\citep{punn2021rca} are considered to establish better comparative analysis of the segmentation performance. In contrast to the U-Net model, A-Unet adds attention filters in the skip connection to suppress irrelevant features of an input image, while following a similar encoder-decoder structure. Later, to efficiently capture the varied shape, size and location of the target structure, I-Unet introduces inception convolution layers where multi-scale features are extracted at the same layer, thereby forming a wider network. Moreover, a hybrid pooling layer is proposed that exploits the features of spatial max pooling and spectral pooling. Inspired from the potential of A-Unet and I-Unet, the RCA-IUnet model advances the attention filter to capture multi-scale feature maps and generate better attention descriptors for target regions, while also using the hybrid pooling and inception convolution layers by reducing the cost of computation and training parameters with the help of depthwise separable convolution~\citep{chollet2017xception}.
	
	\section{Proposed framework}\label{sec4}
\begin{figure}[H]
		\centering
		\includegraphics[width=\columnwidth] {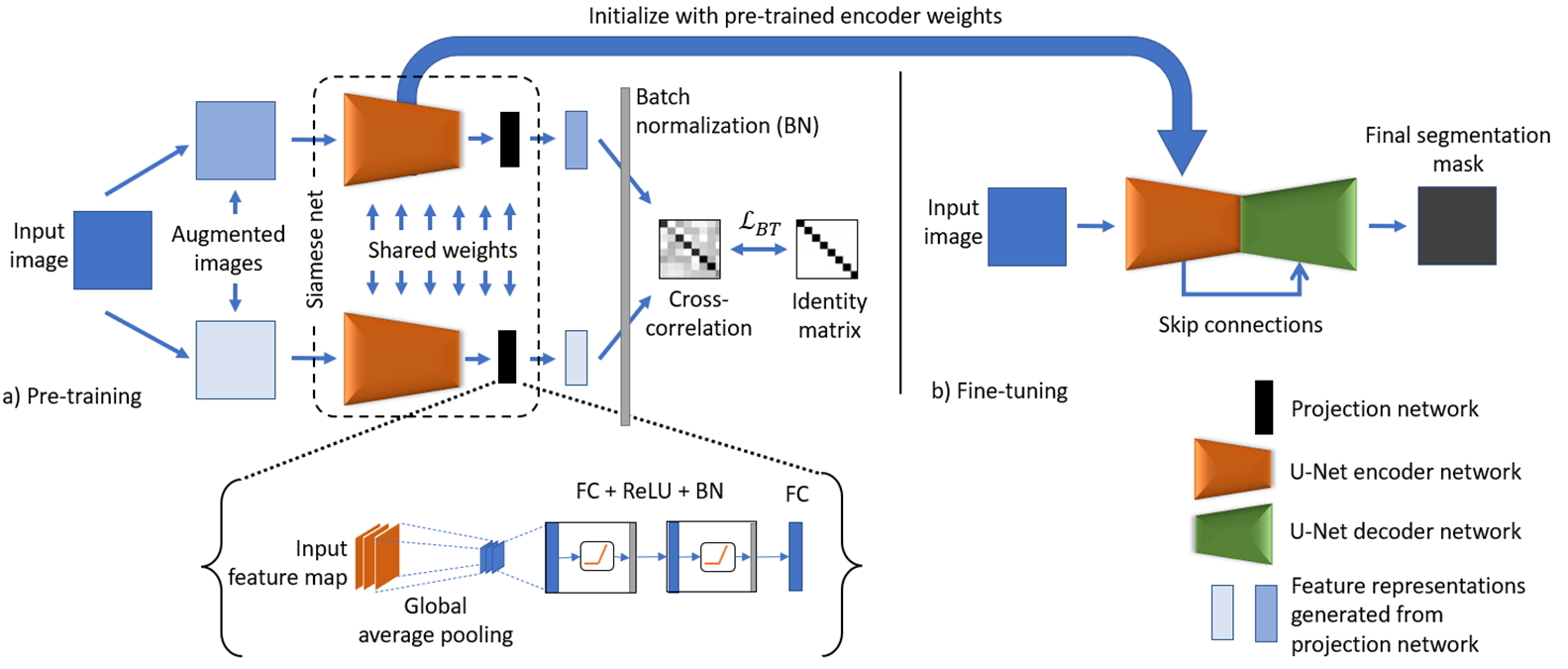}
		\caption{BT-Unet framework. a) Pre-training U-Net encoder network, and b) Fine-tuning U-Net model that is initialized with pre-trained encoder weights.}
		\label{fig3}
	\end{figure}

	In the present article, the state-of-the-art potential of U-Net models is expanded by integrating redundancy reduction based Barlow Twins self-supervised learning for better performance in the segmentation with limited annotated data samples. The schematic representation of the proposed framework is shown in Fig.~\ref{fig3}.
	
	The BT-Unet framework is divided into two phases: 1) Pre-training, and 2) Fine-tuning. In pre-training, the aim is to learn the complex feature representations using unannotated data samples. Here, the encoder network of the U-Net model is pre-trained using BT self-supervised learning strategy. Initially, the input image is augmented or corrupted with certain distortions such as random crop and rotations to generate two distorted images. This type of distortion follows from the results acquired by \cite{zbontar2021barlow} while analyzing the effect of applying augmentations on pre-training performance. Each augmented image is analyzed with a U-Net encoder followed by a projection network to generate encoded feature representations in desired dimensions. The projection network follows from the feature maps produced by the encoder network with global average pooling and blocks of fully connected layers, ReLU activation and batch normalization (FC + ReLU + BN), and final encoded feature representations are generated by another FC layer. Following from the empirical observations, the number of neurons in each fully connected layer is kept half of the spatial dimension of an input image for efficient pre-training, e.g., if input, $I\in \mathbb{R}^{s\times{s} \times{c}} $ then the number of neurons are $s/2$, where $s$ is a spatial dimension of an image. The number of neurons could be further increased but at the cost of heavy computation. However, no significant improvement was observed with increased dimensions.	Since, in later layers, the network learns task specific features that are not aligned with the downstream segmentation task, hence the weights learned by the projection network can be neglected, whereas the weights of the entire encoder network can be transferred to the U-Net model. In the second phase, the weights of the encoder network in the U-Net model are initialized with pre-trained weights (from the first phase), whereas the rest of the network is initialized with default weights. Finally, the U-Net model is fine-tuned with limited annotated samples for the biomedical image segmentation. 
	
	\section{Experiment configuration}\label{sec5}
	This section covers the details concerning the training and testing environment of the BT-Unet framework along with the datasets and modifications in U-Net models that are used for the comparative analysis. To establish robust results with the BT-Unet framework various state-of-the-art U-Net models are considered for experiments such as vanilla U-Net, attention U-Net (A-Unet), inception U-Net (I-Unet) and residual cross-spatial attention guided inception U-Net (RCA-IUnet). Inspired by the RCA-IUnet model, the following minor modifications for U-Net, A-Unet and I-Unet architectures are performed:
	\begin{itemize}
		\item Standard 2D convolution operations are replaced with 2D depthwise separable convolution to reduce the number of training parameters and multiplication operations without affecting performance.
		\item Batch normalization is performed after every convolution operation for stable training.
		\item Each encoder layer is equipped with residual skip connection (mini-skip connection) to avoid the vanishing gradient problem.
		\item Encoding and decoding phases are divided into four stages. With each stage in the encoding phase, the number of channels increases by a factor of 2 (starting with 16 in the first layer) and spatial resolution decreases by a factor 2 (starting with 256$\times$256).
	\end{itemize}
	
	\subsection{Dataset description and setup}
	The performance of the BT-Unet framework is validated using four datasets with different segmentation tasks as shown in Table~\ref{tab1}. The dataset comprises images of organs, cells and lesions acquired under different imaging protocols.
	
	\begin{table}[!t]
		\centering
		\caption{Summary of biomedical datasets used in our experiment.}
		\label{tab1}
		\begin{tabular}{lp{1.5in}p{0.3in}p{0.7in}} 
			\toprule
			Dataset      & Description                                               & Images & Size  \\ 
			\midrule
			KDSB18 \citep{kaggle}           & Nuclei segmentation using
			histopathological cell images.            & 670                     & 256$\times$256        \\
			BUSIS \citep{xian2018benchmark} & Breast tumor segmentation using ultrasound
			scans.                   & 562                     & Variable         \\ 
			ISIC18 \citep{isic18}           & Skin lesion segmentation using dermoscopy
			images.                   & 2596                    & Variable         \\ 
			BraTS18 \citep{brats18}         & Brain tumor segmentation using MRI
			modalities (T1, T1C, T2, FLAIR). & 285                     & 240$\times$240$\times$155  \\
			\botrule
		\end{tabular}
	\end{table}
	
	The Kaggle data science bowl 2018 (KDSB18) challenge is developed for automated nuclei segmentation. It contains annotated histopathological images with varying nuclei shapes, cell types, magnification and modalities (fluorescence/brightfield). Breast ultrasound image segmentation (BUSIS) benchmark dataset comprises breast ultrasound scans annotated with a binary mask of the tumor regions. The dataset covers a wide diversity of samples collected from various medical institutes and organizations. In another dataset ISIC18, skin lesion segmentation is performed with the help of annotated dermoscopy images. To add more diversity in the datasets, brain tumor segmentation 2018 (BraTS18) challenge is considered, which comprises of 3D volumes of MRI modalities with high-grade gliomas (HGG) and low-grade gliomas (LGG) to highlight different tumor regions: whole tumor (WT), tumor core (TC), and emerging tumor (ET). This task is simplified by extracting 4,200 2D slices from the 3D volumes of FLAIR modality and analyzing the segmentation mask associated with the WT region.
	
	\subsection{Training and testing}
	The overall framework is developed using the TensorFlow v2.6 library on Nvidia RTX 2070 Max-Q GPU system. For all experiments, images from KDSB18, BUSI, ISIC18 and BraTS18 datasets are resized to 256$\times$256. The datasets are split into $70\%$ training data and $30\%$ testing data. The pre-training is performed with complete training data without considering annotations. To simulate the scenario of limited annotated data availability, $20\%$ of KDSB18 and BUSI training data and $10\%$ of ISIC18 and BraTS18 training data are considered for fine-tuning the segmentation models with 5-fold and 10-fold cross-validation respectively. Moreover, Adam optimizer with learning rate initialized at $1e-3$ is used for all the experiments that decay by a factor of 0.1 once the learning stagnates for better segmentation results. The training phase is also assisted with the early-stopping strategy to avoid the overfitting problem by terminating the training process when the loss function stops decreasing. The pre-training is performed by minimizing the cross-correlation loss function defined in Eq.~\ref{eq1}, whereas U-Net models are fine-tuned with segmentation loss function, $\mathcal{L}$ defined as the average of binary cross-entropy loss, $\mathcal{L}_{BC}$ and dice coefficient loss, $\mathcal{L}_{DC}$ as shown in Eq.~\ref{eq4}.
	\begin{equation}
		\mathcal{L}=\frac{1}{2}    \mathcal{L}_{BC}+\frac{1}{2}\mathcal{L}_{DC}
		\label{eq4}
	\end{equation}
	\begin{equation}
		\begin{aligned}
			\mathcal{L}_{BC}\left(y,p\left(y\right)\right)=-\sum^N_i\left(y_i.{log \left(p\left(y_i\right)\right)}+\left(1-y_i\right).{log \left(1-p\left(y_i\right)\right)}\right)
		\end{aligned}
		\label{eq5}
	\end{equation}
	\begin{equation}
		\mathcal{L}_{DC}\left(y,p\left(y\right)\right)=1-\frac{2\sum^N_i{y_i.p(y_i)}}{\sum^N_i\mid y_i\mid ^2+\sum^N_i\mid p(y_i)\mid ^2}
		\label{eq6}
	\end{equation}
	where $y$ is the ground truth label of a pixel, $p(y)$ is the predicted label of a pixel and $N$ is the total number of pixels.
	
	The performance of trained U-Net models is validated on the test sets by using various evaluation metrics such as precision (Pr.) (Eq.~\ref{eq7}), dice coefficient (DC) (Eq.~\ref{eq8}) and mean intersection-over-union (mIoU) (Eq.~\ref{eq9}).
	\begin{equation}
		Pr. = \frac{TP}{(TP+FP)}
		\label{eq7}
	\end{equation}
	\begin{equation}
		DC = \frac{2TP}{(2TP+FN+FP)}
		\label{eq8}
	\end{equation}
	\begin{equation}
		mIoU = \frac{1}{10}\sum_t\frac{TP}{(TP+FN+FP)};\;\;\; t+=0.5\leq 0.95
		\label{eq9}
	\end{equation}
	where, $TP$ - true positive, $TN$ - true negative, $FP$ - false positive, $FN$ - false negative and $t$ - prediction threshold.
	
	\section{Results and discussion}\label{sec6}
\begin{figure}[!t]
		\centering
		\includegraphics[width=\columnwidth] {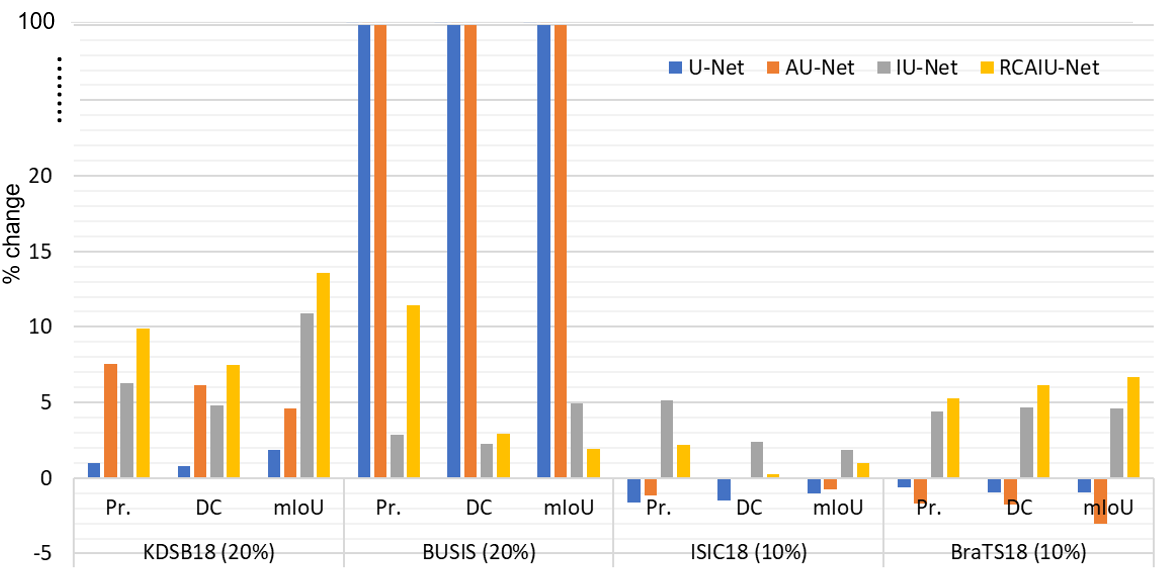}
		\caption{Impact of BT pre-training on segmentation performance of the U-Net models.}
		\label{fig4}
	\end{figure}

\begin{figure}[!t]
		\centering
		\includegraphics[width=\columnwidth] {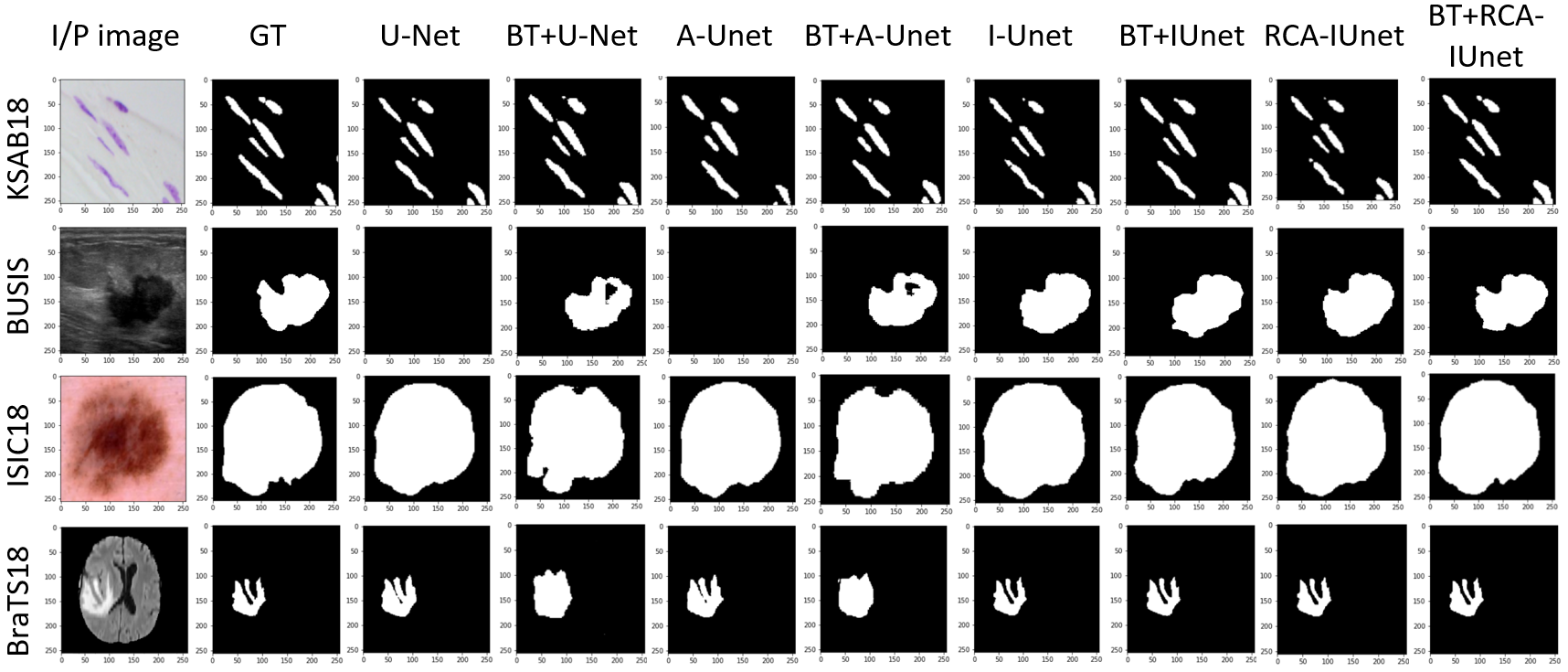}
		\caption{Qualitative comparative analysis of the segmentation performance.}
		\label{fig5}
	\end{figure}

	The proposed framework generates a segmentation mask for given medical imaging. The quantitative results of the U-Net models with and without the Barlow Twins based pre-training on four different biomedical imaging datasets is presented in Table~\ref{tab2} along with the percentage change in the segmentation performance of the models with Fig.~\ref{fig4}. Moreover, Fig. \ref{fig5} presents the qualitative comparison of the segmentation performance. Following are the observations made for each dataset:
	
	\begin{table}[!b]
		\centering
		\caption{Quantitative comparative analysis of segmentation models. The best results with and without BT are shown with bold black and blue color respectively.}
		\label{tab2}
		\resizebox{\textwidth}{!}{\begin{tabular}{p{0.3in}lp{0.2in}p{0.2in}lp{0.2in}p{0.2in}lp{0.2in}p{0.2in}lp{0.2in}p{0.2in}p{0.2in}} 
				\toprule
				\multirow{2}{*}{{Model}} & \multirow{2}{*}{BT} & \multicolumn{3}{c}{KDSB18 (20\%)}                                                              & \multicolumn{3}{c}{BUSIS (20\%)}                                                               & \multicolumn{3}{c}{ISIC18 (10\%)}                                                               & \multicolumn{3}{c}{BraTS18 (10\%)}                                                               \\ 
				\cmidrule{3-14}
				&   & Pr. & DC & mIoU & Pr. & DC & mIoU & Pr. & DC & mIoU & Pr. & DC & mIoU  \\ 
				\midrule
				\multirow{2}{*}{U-Net}                         & N                                                   & \textcolor{blue}{\textbf{0.904}}  & \textcolor{blue}{\textbf{0.911}} & \textcolor{blue}{\textbf{0.891}}  & 0.000                             & 0.000                            & 0.000                             & 0.894                             & 0.803                            & 0.797                             & 0.518                             & 0.543                            & 0.423                              \\ 
				%\cline{2-14}
				& Y                                                   & 0.913                             & 0.918                            & 0.912                             & 0.823                             & 0.676                            & 0.547                             & 0.880                             & 0.791                            & 0.789                             & 0.515                             & 0.538                            & 0.419                              \\ 
				%\hline
				\multirow{2}{*}{A-Unet}                        & N                                                   &  0.811                            & 0.824                            & 0.788                             & 0.000                             & 0.000                            & 0.000                             & 0.885                             & 0.811                            & 0.806                             & 0.600                             & 0.573                            & 0.543                              \\ 
				%\cline{2-14}
				& Y                                                   & 0.877                             & 0.878                            & 0.826                             & 0.662                             & 0.639                            & 0.597                             & 0.875                             & 0.811                            & 0.800                             & 0.590                             & 0.563                            & 0.527                              \\ 
				%\hline
				\multirow{2}{*}{I-Unet}                        & N                                                   &  0.865                            & 0.873                            & 0.809                             & 0.789  & 0.738                            & 0.731                             & 0.871                             & 0.813                            & 0.804                             & 0.622                             & 0.599                            & 0.537                              \\ 
				%\cline{2-14}
				& Y                                                   & 0.923                             & 0.917                            & 0.908                             & 0.771                             & 0.769                            & 0.807                             & 0.903                             & 0.830                            & 0.819                             & 0.650                             & 0.628                            & 0.563                              \\ 
				%\hline
				\multirow{2}{0.3in}{RCA-IUnet}                     & N                                                   &  0.839                            & 0.852                            & 0.789                             & \textcolor{blue}{\textbf{0.794}}   & \textcolor{blue}{\textbf{0.748}} & \textcolor{blue}{\textbf{0.756}}  & \textcolor{blue}{\textbf{0.883}}  & \textcolor{blue}{\textbf{0.828}} & \textcolor{blue}{\textbf{0.813}}  & \textcolor{blue}{\textbf{0.689}}  & \textcolor{blue}{\textbf{0.664}} & \textcolor{blue}{\textbf{0.628}}   \\ 
				%\cline{2-14}
				& Y                                                   & \textbf{0.931}                    & \textbf{0.921}                   & \textbf{0.913}                    & \textbf{0.833}                    & \textbf{0.818}                   & \textbf{0.771}                    & \textbf{0.918}                    & \textbf{0.833}                   & \textbf{0.821}                    & \textbf{0.734}                    & \textbf{0.701}                   & \textbf{0.673}                     \\
				\botrule
		\end{tabular}}
	\end{table}
	
	\begin{itemize}
		\item \textbf{KDSB18}. In the cell nuclei segmentation task of KDSB18, the performance of the BT enabled U-Net models exceeds as compared to the models without BT (as shown in Table~\ref{tab2}). It is also observed that as the architecture design becomes more complex then pre-training exhibits positive influence on the segmentation performance (as shown in Fig.~\ref{fig4}, RCA-IUnet model achieves maximum gain in the performance as compared to other models). Moreover, the minimum change in the performance of the vanilla U-Net model indicates that a simpler encoder structure (close to vanilla U-Net, e.g. A-Unet) face difficulty in extracting feature maps with limited annotated samples. A similar pattern can also be observed with qualitative results in Fig.~\ref{fig5}.
		
		\item \textbf{BUSIS}. The automated segmentation of breast tumor using ultrasound imaging achieves promising results with BT pre-training (as shown in Table~\ref{tab2}). It is observed that U-Net and A-Unet models are not able to learn and extract feature maps concerning tumor regions (achieved 0 precision, DC and mIoU), however with pre-training, these models achieved noticeable improvement (as shown in Fig.~\ref{fig5}). In case of I-Unet and RCA-IUnet models, considerable improvements are observed with pre-training, where dice coefficient increases by $5\%$ and precision increases by $11\%$ respectively (as shown in Fig.~\ref{fig4}).
		
		\item \textbf{ISIC18}. Skin lesion segmentation is another challenging task, where U-Net models with BT achieved satisfactory improvements in segmentation. The I-Unet and RCA-IUnet models are the most influenced networks that achieved $5.1\%$ and $2.2\%$ increase in precision respectively. However, a slight decline in performance is observed with vanilla U-Net and A-Unet (as shown in Fig.~\ref{fig4}) while using BT pre-training. In contrast to I-Unet and RCA-IUnet, these models have simpler encoder structures due to which in pre-training the model fails to learn efficient feature representation about complex lesion regions. Furthermore, as observed from Fig.~\ref{fig5}, the BT+RCA-IUnet model achieved best skin lesion segmentation results with smoother boundaries. 
		
		\item \textbf{BraTS18}. In this challenge of brain tumor segmentation, the models performed similarly as with the ISIC18 dataset. I-Unet and RCA-IUnet models achieved significant gain in the segmentation performance while using the BT-Unet framework, whereas the same behaviour is not observed with vanilla U-Net and A-Unet models because of their inability to effectively capture tumor feature representations during pre-training. As observed from Fig.~\ref{fig4}, the RCA-IUnet model achieved gains of  $5.3\%$, $6.1\%$ and $6.7\%$, while I-Unet achieved gains of $4.4\%$, $4.7\%$ and $4.6\%$ in precision, dice coefficient and mIoU respectively. 
	\end{itemize}
	
	\begin{table}[!b]
		\centering
		\caption{Performance analysis of U-Net variants with and without pre-training using Barlow Twins (BT) over different fractions of training datasets (DS).}
		\label{tab3}
		\resizebox{\textwidth}{!}{\begin{tabular}{lllp{0.2in}p{0.2in}lp{0.2in}p{0.2in}lp{0.2in}p{0.2in}lp{0.2in}p{0.2in}p{0.2in}}
				\toprule
				\multirow{2}{*}{DS} & \multirow{2}{*}{Model}     & \multirow{2}{*}{BT} & \multicolumn{3}{c}{30\%} & \multicolumn{3}{c}{50\%} & \multicolumn{3}{c}{80\%} & \multicolumn{3}{c}{100\%} \\
				\cmidrule{4-15}
				&                            &                     & Pr.    & DC     & mIoU   & Pr.    & DC     & mIoU   & Pr.    & DC     & mIoU   & Pr.     & DC     & mIoU   \\
				\midrule
				\multirow{8}{*}{\rotatebox[origin=c]{90}{KDSB18}}  & \multirow{2}{*}{U-Net}     & N                   & 0.903  & 0.944  & 0.921  & 0.952  & 0.965  & 0.951  & 0.955  & 0.968  & 0.954  & 0.957   & 0.969  & 0.954  \\
				&                            & Y                   & 0.923  & 0.951  & 0.940  & 0.955  & 0.969  & 0.951  & 0.958  & 0.970  & 0.957  & 0.959   & 0.970  & 0.955  \\
				& \multirow{2}{*}{A-Unet}    & N                   & 0.868  & 0.871  & 0.866  & 0.889  & 0.900  & 0.877  & 0.923  & 0.943  & 0.921  & 0.955   & 0.965  & 0.953  \\
				&                            & Y                   & 0.901  & 0.921  & 0.893  & 0.923  & 0.944  & 0.913  & 0.944  & 0.968  & 0.943  & 0.954   & 0.965  & 0.952  \\
				& \multirow{2}{*}{I-Unet}    & N                   & 0.896  & 0.908  & 0.894  & 0.946  & 0.958  & 0.943  & 0.955  & 0.969  & 0.955  & 0.958   & 0.971  & 0.955  \\
				&                            & Y                   & 0.934  & 0.953  & 0.931  & 0.957  & 0.971  & 0.952  & 0.961  & 0.972  & 0.960  & 0.961   & 0.973  & 0.956  \\
				& \multirow{2}{0.3in}{RCA-IUnet} & N                   & 0.873  & 0.888  & 0.865  & 0.950  & 0.964  & 0.948  & 0.964  & 0.970  & 0.963  & 0.966   & 0.971  & 0.964  \\
				&                            & Y                   & 0.944  & 0.900  & 0.899  & 0.971  & 0.972  & 0.970  & 0.980  & 0.981  & 0.977  & 0.967   & 0.972  & 0.965  \\
				\midrule
				\multirow{8}{*}{\rotatebox[origin=c]{90}{BUSIS}}   & \multirow{2}{*}{U-Net}     & N                   & 0.430  & 0.320  & 0.210  & 0.819  & 0.845  & 0.816  & 0.921  & 0.874  & 0.858  & 0.933   & 0.893  & 0.866  \\
				&                            & Y                   & 0.856  & 0.723  & 0.689  & 0.919  & 0.888  & 0.843  & 0.925  & 0.879  & 0.867  & 0.934   & 0.894  & 0.867  \\
				& \multirow{2}{*}{A-Unet}    & N                   & 0.394  & 0.312  & 0.191  & 0.851  & 0.824  & 0.821  & 0.933  & 0.887  & 0.869  & 0.942   & 0.911  & 0.897  \\
				&                            & Y                   & 0.866  & 0.756  & 0.700  & 0.910  & 0.882  & 0.842  & 0.934  & 0.887  & 0.868  & 0.943   & 0.912  & 0.897  \\
				& \multirow{2}{*}{I-Unet}    & N                   & 0.845  & 0.823  & 0.793  & 0.887  & 0.844  & 0.839  & 0.943  & 0.893  & 0.873  & 0.949   & 0.913  & 0.900  \\
				&                            & Y                   & 0.886  & 0.855  & 0.843  & 0.923  & 0.901  & 0.889  & 0.946  & 0.911  & 0.899  & 0.951   & 0.928  & 0.911  \\
				& \multirow{2}{0.3in}{RCA-IUnet} & N                   & 0.881  & 0.861  & 0.823  & 0.908  & 0.889  & 0.869  & 0.934  & 0.901  & 0.888  & 0.954   & 0.937  & 0.921  \\
				&                            & Y                   & 0.911  & 0.873  & 0.855  & 0.923  & 0.901  & 0.898  & 0.952  & 0.926  & 0.911  & 0.956   & 0.938  & 0.922  \\
				\midrule
				\multirow{8}{*}{\rotatebox[origin=c]{90}{ISIC18}}  & \multirow{2}{*}{U-Net}     & N                   & 0.929  & 0.874  & 0.856  & 0.949  & 0.899  & 0.891  & 0.954  & 0.922  & 0.911  & 0.972   & 0.958  & 0.946  \\
				&                            & Y                   & 0.911  & 0.871  & 0.849  & 0.955  & 0.894  & 0.888  & 0.944  & 0.915  & 0.898  & 0.969   & 0.957  & 0.945  \\
				& \multirow{2}{*}{A-Unet}    & N                   & 0.916  & 0.865  & 0.850  & 0.944  & 0.896  & 0.889  & 0.952  & 0.918  & 0.908  & 0.971   & 0.958  & 0.945  \\
				&                            & Y                   & 0.913  & 0.851  & 0.842  & 0.942  & 0.872  & 0.866  & 0.959  & 0.938  & 0.930  & 0.969   & 0.953  & 0.944  \\
				& \multirow{2}{*}{I-Unet}    & N                   & 0.894  & 0.855  & 0.843  & 0.941  & 0.882  & 0.879  & 0.952  & 0.932  & 0.931  & 0.980   & 0.969  & 0.954  \\
				&                            & Y                   & 0.913  & 0.860  & 0.852  & 0.949  & 0.889  & 0.884  & 0.963  & 0.944  & 0.933  & 0.978   & 0.968  & 0.952  \\
				& \multirow{2}{0.3in}{RCA-IUnet} & N                   & 0.893  & 0.855  & 0.843  & 0.948  & 0.883  & 0.880  & 0.964  & 0.943  & 0.934  & 0.982   & 0.969  & 0.959  \\
				&                            & Y                   & 0.939  & 0.869  & 0.863  & 0.969  & 0.919  & 0.913  & 0.969  & 0.959  & 0.943  & 0.981   & 0.968  & 0.958  \\
				\midrule
				\multirow{8}{*}{\rotatebox[origin=c]{90}{BraTS18}} & \multirow{2}{*}{U-Net}     & N                   & 0.555  & 0.551  & 0.438  & 0.601  & 0.589  & 0.532  & 0.653  & 0.623  & 0.615  & 0.652   & 0.623  & 0.614  \\
				&                            & Y                   & 0.549  & 0.542  & 0.449  & 0.588  & 0.574  & 0.528  & 0.644  & 0.620  & 0.609  & 0.651   & 0.622  & 0.613  \\
				& \multirow{2}{*}{A-Unet}    & N                   & 0.681  & 0.652  & 0.621  & 0.723  & 0.689  & 0.676  & 0.761  & 0.743  & 0.721  & 0.769   & 0.755  & 0.723  \\
				&                            & Y                   & 0.653  & 0.622  & 0.599  & 0.713  & 0.677  & 0.666  & 0.758  & 0.741  & 0.719  & 0.765   & 0.752  & 0.721  \\
				& \multirow{2}{*}{I-Unet}    & N                   & 0.690  & 0.673  & 0.620  & 0.752  & 0.711  & 0.699  & 0.798  & 0.763  & 0.754  & 0.844   & 0.813  & 0.808  \\
				&                            & Y                   & 0.721  & 0.705  & 0.661  & 0.773  & 0.733  & 0.724  & 0.813  & 0.792  & 0.760  & 0.860   & 0.821  & 0.818  \\
				& \multirow{2}{0.3in}{RCA-IUnet} & N                   & 0.773  & 0.734  & 0.713  & 0.822  & 0.788  & 0.762  & 0.864  & 0.834  & 0.811  & 0.923   & 0.899  & 0.875  \\
				&                            & Y                   & 0.811  & 0.778  & 0.763  & 0.842  & 0.797  & 0.772  & 0.882  & 0.845  & 0.839  & 0.925   & 0.901  & 0.878 \\
				\botrule
			\end{tabular}}
	\end{table}
	
	Furthermore, the performance of U-Net variants with and without the pre-training is analysed with multiple fractions of training datasets as shown in Table~\ref{tab3}. For all datasets with training fractions less than $50\%$, similar change in performance is observed among the models as discussed in Table~\ref{tab2}. Besides, in the case of without pre-training for U-Net and A-Unet, the increased fraction of BUSIS training samples improved the performance as compared to zero values (observed in Table~\ref{tab2}) and the corresponding performance gain is also achieved with pre-training. However, for the fractions greater than $50\%$, the performance gap is narrowed i.e. results produced by the models with and without pre-training are not significantly different. This indicates that it is beneficial to utilize the pre-training strategy when there are limited annotations within the large pool of data samples.
	
	\section{Conclusion}\label{sec7}
	In this research work, self-supervised learning assisted biomedical image segmentation framework BT-Unet, is proposed to address one of the major challenges of limited annotated data availability. The BT-Unet framework uses redundancy reduction based Barlow Twins strategy for pre-training the encoder network of the U-Net model with feature representations of the data in an unsupervised manner, followed by fine-tuning of the U-Net model for downstream biomedical image segmentation task with limited annotated data samples. With exhaustive experimental trials, it is evident that BT-Unet tends to improve the segmentation performance of U-Net models in such situations. Moreover, this improvement is also influenced by the underlying encoder structure and nature of biomedical image segmentation task. In future, more experiments can be conducted by modifying or exploring different pre-training strategies to generate better feature representations and ensure finer biomedical image segmentation.

	\section*{Declarations}
	
	\begin{itemize}
		\item Acknowledgment: We thank our institute, Indian Institute of Information Technology Allahabad (IIITA), India and Big Data Analytics (BDA) lab for allocating necessary resources to perform this research. We extend our thanks to our colleagues for their valuable guidance and suggestions.
		\item Conflict of interest: The authors have no relevant financial or non-financial interests to disclose.
		\item Ethics approval: Not Applicable.
		\item Consent to participate: Not Applicable.
		\item Consent for publication: Not Applicable.
		\item Authors' contributions: All authors contributed equally in conceptualizing the research problem and preparation of the manuscript.
		\item Availability of data: All datasets are publicly accessible.
		\item Code availability: Code for using BT-Unet framework is available at \url{https://github.com/nspunn1993/BT-Unet}.
	\end{itemize}

	\bibliography{reference}% common bib file
	
	
\end{document}